\documentclass[a4paper]{article}
\usepackage{interspeech2011,amssymb,amsmath,epsfig,booktabs,tipa,array,multirow,url}
\usepackage{url}
\ninept
\def\reg{{\rm\ooalign{\hfil
     \raise.07ex\hbox{\scriptsize R}\hfil\crcr\mathhexbox20D}}}

\title{Audio and Contact Microphones for Cough Detection}

\makeatletter
\def\name#1{\gdef\@name{#1\\}}
\makeatother
\name{{\em Thomas Drugman$^1$, Jerome Urbain$^1$, Nathalie Bauwens$^2$,} \\
{\em Ricardo Chessini$^1$, Anne-Sophie Aubriot$^2$, Patrick Lebecque$^2$, Thierry Dutoit$^1$}}

\address{$^1$TCTS Lab - University of Mons, Belgium \\ $^2$Pediatric Pulmonology \& Cystic Fibrosis Unit, Cliniques St Luc, University of Louvain, Belgium\\
  {\small \tt thomas.drugman@umons.ac.be}}

\begin{document}
\maketitle

\begin{abstract}
In the framework of assessing the pathology severity in chronic cough diseases, medical literature underlines the lack of tools for allowing the automatic, objective and reliable detection of cough events. This paper describes a system based on two microphones which we developed for this purpose. The proposed approach relies on a large variety of audio descriptors, an efficient algorithm of feature selection based on their mutual information and the use of artificial neural networks. First, the possible use of a contact microphone (placed on the patient's thorax or trachea) in complement to the audio signal is investigated. This study underlines that this contact microphone suffers from reliability issues, and conveys little new relevant information compared to the audio modality. Secondly, the proposed audio-only approach is compared to a commercially available system using four sensors on a database with different sound categories often misdetected as coughs, and produced in various conditions. With average sensitivity and specificity of 94.7\% and 95\% respectively, the proposed method achieves better cough detection performance than the commercial system.
\end{abstract}

\section{Introduction}\label{sec:Intro}
Cough is the commonest reason for which patients seek medical advice to the general practitioner (around 20\% of consultations for children below 4 years old), the paediatrician and the pneumologist (for whom chronic cough represents one third of consultations). The impact of cough, notably chronic coughing, on life quality can be important.

In order to evaluate the cough severity, a subjective assessment is possible by making use of cough diaries, quality-of-life questionnaires or relying on a visual analog scale. However, it has been shown that the subjective perception of cough is only slightly correlated with objective measurements of its severity \cite{Decalmer}. Medical literature on this topic therefore underlines the lack of a tool allowing the automatic, objective and reliable quantification of this symptom \cite{ERJ}. This latter step is notably required prior to any correct evaluation of possible treatments.

Some approaches have been recently proposed to address the automatic detection of cough \cite{Smith}. These systems generally couple various sensors to the audio signal \cite{Smith}: accelerometer, chest impedance belt, contact microphone, ECG, respiratory inductance plethysmography etc. Although reported results are encouraging \cite{Smith}, there is currently neither standardized methods nor adequately validated, commercially available and clinically acceptable cough monitors. Besides, following the patient in ambulatory and 24h-long conditions (while preserving his daily habits) remains an open problem. As a result, cough quantification in the majority of hospitals is still nowadays performed by a tedious task of manual counting from audio recordings, or for validation by comparison using simultaneous video recordings.

\begin{figure}[h]
  \begin{center}
   \includegraphics[width=8cm]{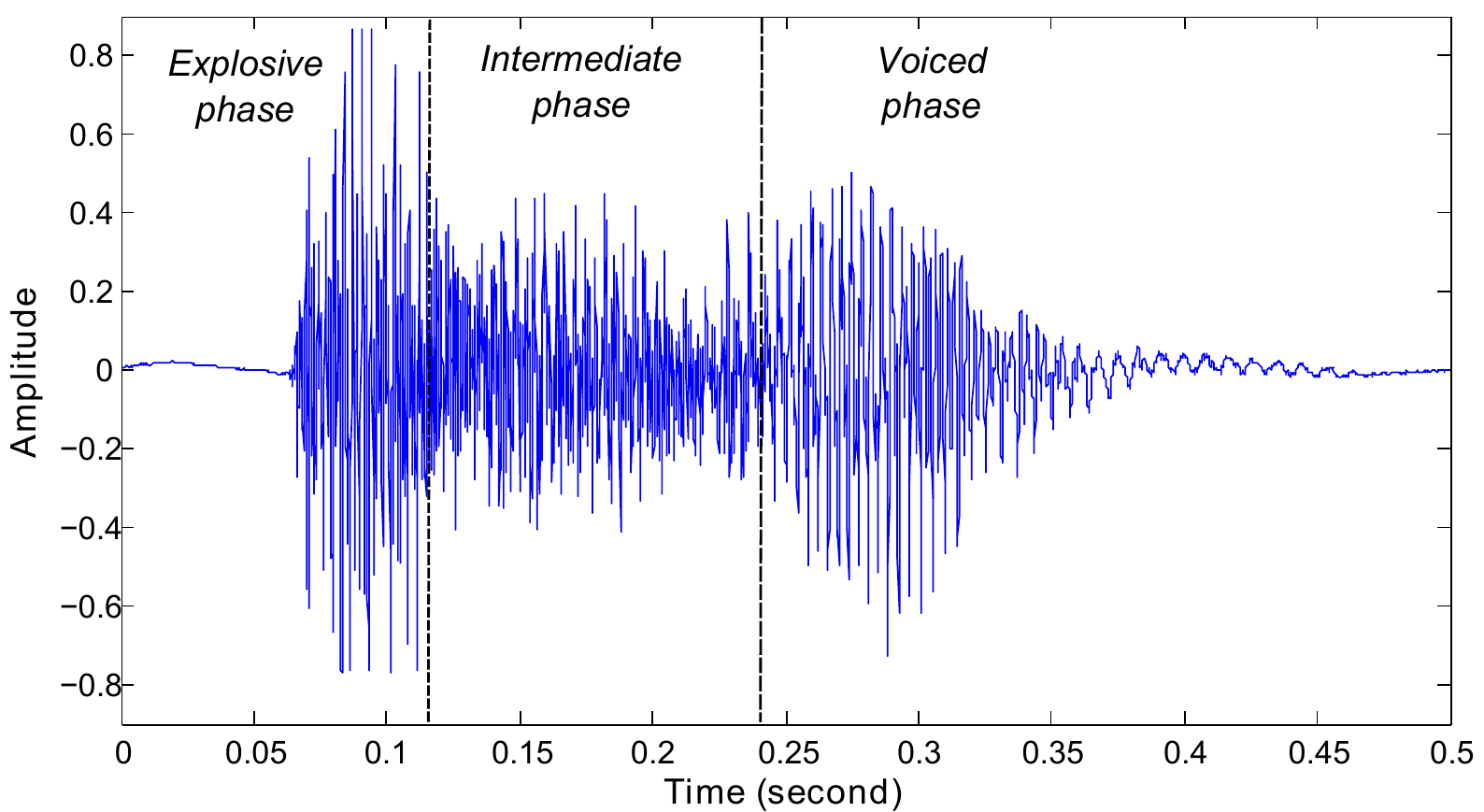}
  \end{center}
\vspace{-.5cm}
\caption{{\it Waveform of a typical cough sound with three phases.}} 
\vspace{-.2cm}
\label{fig:Example}
\end{figure}

For respiratory physiologists, cough is three-phase expulsive motor act characterized by an inspiratory effort, followed by a forced expiratory effort against a closed glottis and then by opening of the glottis and rapid expiratory airflow \cite{ERJ}. As shown in Figure \ref{fig:Example}, the acoustics of the cough sound is manifested by three phases, where the last one is optional \cite{Korpas}: an explosive phase, an intermediate period whose characteristics are similar to a forced expiration, and a voiced phase.

The goal of this paper is to develop an automatic cough counter based on the audio signal and possibly an additional contact microphone placed on the patient's thorax or trachea. For this, the structure of this article is the following. Section \ref{sec:Database} describes the database which is used throughout the rest of the paper. The framework we opted for is then detailed in Section \ref{sec:Framework}. The contribution brought by the contact microphone placed on the thorax or on the trachea is investigated in Section \ref{sec:Thorax}. The proposed audio-based cough detector is then compared in Section \ref{sec:KS} to an existing device, the Karmelsonix system \cite{Vizel}. Finally Section \ref{sec:conclu} concludes the paper.

\section{Database}\label{sec:Database}
The key idea of the database was to involve, in various environmental conditions, an important number of subjects producing not only several types of cough events, but also other typical sounds (such as expiration, throat clearing, laugh or speech) for which confusion with cough can be high. The corpus was acquired on healthy people using an audio microphone and a contact microphone. It consists of two separate sets of recordings sampled at 10kHz. Set A was recorded on 22 subjects for which the contact microphone was placed on the trachea. In set B, this latter sensor was placed on the thorax of 10 other subjects and parallel recordings with the Karmelsonix system were also acquired. Karmelsonix is a commercially available cough counter briefly described in Section \ref{sec:KS}.

For each subject from both sets, three sessions were recorded. Each session followed the same protocol but was carried out in three different conditions: sitting down, sitting down with an ambient noise (TV program in the vicinity of the subject) and going up/down stairs. For each session, subjects were asked to produce cough in its diversity: 5 cough events with high volume, 5 with middle and 5 with low volume, and 3 fits of coughing containing 3 events each. Besides they were also asked to produce \emph{parasitical} sounds where there could be an ambiguity for the detection: 3 forced expirations, 5 throat clearings, speech (about 20 seconds) and 3 laugh events. The total database then contains (although slight deviations to the protocol were observed) about 2304 cough events, 288 forced expirations, 480 throat clearing, 32 minutes of speech and 288 laughs, for a total duration of 4 hours. All these audio events were manually annotated on the whole corpus.

\section{Proposed Framework}\label{sec:Framework}

\begin{figure}[h]
  \begin{center}
   \includegraphics[width=8cm]{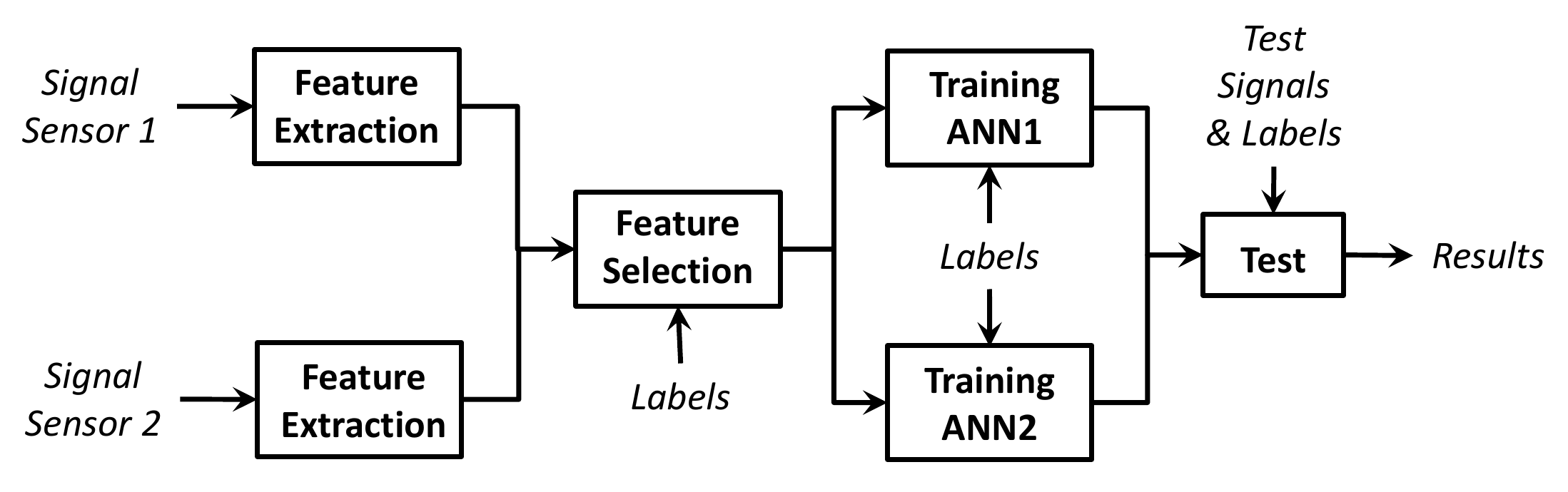}
  \end{center}
\vspace{-.5cm}
\caption{{\it Workflow of the proposed approach.}} 
\vspace{-.2cm}
\label{fig:Workflow}
\end{figure}

The workflow of the proposed approach is displayed in Figure \ref{fig:Workflow}. From the signals possibly captured by several sensors (in our case, the audio and contact microphones), the first step extracts a wide variety of features, as described in Section \ref{ssec:Extraction}. Based on some measures derived from the Information Theory, only a limited number of relevant features are selected (Section \ref{ssec:Selection}). Two distinctive Artificial Neural Networks (ANNs) are then trained for different purposes, as explained in Section \ref{ssec:ANN}. Finally, Section \ref{ssec:Test} presents how the trained models are tested and how the system is evaluated.

\subsection{Feature Extraction}\label{ssec:Extraction}

Many audio features have been proposed in the literature, some being more suited for certain applications. The key idea here is to extract the largest variety of features among which only the most relevant will be selected. These features are extracted every 12 ms on a 30ms-long window and can be divided into two categories: features describing the spectral contents and measures of noise. We also added the first and second derivatives for each of these features in order to integrate the sound dynamics.   

\subsubsection{Features Describing the Spectral Contents}\label{ssec:Spectral}
Several features characterizing the spectral shape have been proposed in \cite{Peeters}. For a comprehensive description of the magnitude spectrum, we used the widely-used \textit{MFCCs}, the \textit{loudness} associated to each Bark band \cite{Peeters} and the \textit{relative energy} in different frequency subbands. Besides, several parameters describing the spectral shape are also employed. The \emph{Spectral Centroid} is defined as the barycenter of the amplitude spectrum. Similarly, the \emph{Spectral Spread} is the dispersion of the spectrum around its mean value. The \emph{Spectral Decrease} is a perceptual measure quantifying the amount of decreasing of the spectral amplitude \cite{Peeters}. Finally, the \emph{Spectral Variation} and \emph{Spectral Flux} characterize the amount of variations of spectrum along time and are based on the normalized cross-correlation between two successive amplitude spectra \cite{Peeters}. Besides, we also use the energy and total loudness which are informative mainly about the presence of audio activity.

\subsubsection{Measures of Noise}\label{ssec:Noise}
Quantifying the level of noise in the signal is of interest for describing the cough sound. For this purpose, several measures are here extracted. First, the \emph{Harmonic to Noise Ratio} (HNR) is calculated in four frequency ranges. The \emph{Spectral Flatness} measures the noisiness/sinusoidality of a spectrum (or a part of it) in four frequency bands \cite{Peeters}. The \emph{Zero-Crossing Rate} quantifies the number of times the signal crosses the zero axis. It is expected that the greater the amount of noise, the higher the number of zero-crossings. The $F_0$ value and its related measure of periodicity based on the Summation of Residual Harmonics \cite{SRH} are used as voicing measurements. As a last parameter quantifying the amount of noise in the audio signal, the \emph{Chirp Group Delay} is a phase-based measure proposed in \cite{DrugmanPhase} for highlighting turbulences during glottal production.

\subsection{Feature Selection}\label{ssec:Selection}
A total number of 222 features (including the first and second derivatives) has been extracted in Section \ref{ssec:Extraction}. The goal of the feature selection algorithm is to retain the most relevant ones so as to alleviate the effect of the curse of dimensionality \cite{Bellman}. For this, we here make use of measures based on the mutual information. This allows to assess the intrinsic discrimination power of each feature separately, but also their possible complementarity or redundancy. And this independently of the subsequent classifier.

The algorithm used in this study has been proposed in \cite{Drugman-FS}. It is a greedy method which at each iteration chooses the feature conveying the greatest amount of new relevant information for the considered classification problem. This latter measure is estimated by considering the mutual information of this feature and its redundancy with the selected subset.

In the rest of the paper, 50 features are used for each configuration (i.e possibly combining features from the audio and contact microphones). In general, these features arise from the two categories of descriptors presented in Section \ref{ssec:Extraction}.

\subsection{Training the Artificial Neural Networks}\label{ssec:ANN}
The key idea of our approach is to focus only on the detection of the explosive phase of the cough sound (see Figure \ref{fig:Example}). Indeed the intermediate phase of cough is very similar to a forced expiration. Besides, the voiced phase does not appear in all cough sounds and could also be confused with some parts of laughing or throat clearing. On the opposite, the explosive phase is charateristic of the beginning of any cough sound and is acoustically distinctive enough from possible parasitical signals. Since the explosive phases have not been manually annotated, we consider in the remainder of this paper that they dominate the first 60ms of the cough events. 

Given that explosive phases are very underrepresented in the database (compared to other sounds), we decomposed the problem into two subtasks:
\begin{itemize}

\item removing the segments of background noise and speech (\emph{\{Cough, Forced Expiration, Throat Clearing, Laugh\}$ vs. $\{Background noise, Speech\}}). Here, the energy during the explosive phase is sufficiently high which implies that we can be almost sure that no segment of interest will be removed.

\item discriminating the explosive phases of cough from other ambiguous sounds (\emph{\{Explosive phase\}$ vs. $\{Intermediate phase, Voiced phase, Forced Expiration, Throat Clearing, Laugh, Speech\}}). 


\end{itemize}

For each subtask, a dedicated Artifical Neural Network (ANN) has been trained. Our ANN implementation relies on the Matlab Neural Network toolbox. Each ANN is made of a single hidden layer consisting of neurons (fixed to 32 in this work) whose activation function is an hyperbolic tangent sigmoid transfer function. The output layer is a simple neuron with a logarithmic sigmoid function suited for a binary decision.

\subsection{Testing and Evaluation}\label{ssec:Test}

When new test recordings are provided to the system, the posterior probabilities for the two ANNs are computed. Since the explosive phases are the only segments for which both ANNs should ideally give a unitary output, their posteriors are combined by multiplication. The resulting output is then smoothed by a median filtering over a period of 50ms so as to remove isolated decisions.

In Section \ref{sec:Thorax}, the system is evaluated by a leave-one-out (at the subject level) cross-validation approach. This means that training is achieved on the whole dataset except one subject which is left for the test, and this operation is repeated so as to cover the whole dataset for the evaluation. 

For assessing the performance, we calculate the number of misses and false alarms at the event level, and derive the \emph{specificity} and \emph{sensitivity} metrics.

%
%

\section{Using a Contact Microphone on the Thorax or on the Trachea}\label{sec:Thorax}

It is here investigated whether our contact microphone placed on the patient's trachea or thorax can be useful in complement with the audio signal. In a first time, we focus on set A of the database where the contact microphone was placed on the trachea of 22 subjects. Three versions of our system are considered here: using only the audio microphone, only the contact microphone, or using both together. For each configuration, 50 features were selected. The threshold for the decision was varied so as to have a good compromise between misses and false alarms, i.e so as to find a trade-off between specificity and sensitivity.

\begin{table} 
  \begin{center}
\vspace{0cm}
    \begin{tabular}{c|c|c|}
 & \bf{Specificity (\%)} & \bf{Sensitivity (\%)} \\
\hline 
\hline 
\bf{Audio} & 87.85 & 87.69 \\
\hline 
\bf{Trachea} & 71.37 & 71.71 \\
\hline 
\bf{Audio + Trachea} & 86.36 & 86.73 \\
\hline 
    \end{tabular}
  \end{center}
  \vspace{-0.5cm}
  \caption{\emph{Results of cough event detection using the contact microphone placed on the trachea (set A of the database).}}
  \label{tab:Trachea}
\vspace{-.2cm}
\end{table}

Results are shown in Table \ref{tab:Trachea}. It is observed that the audio microphone provides much better results than the contact microphone placed at the trachea, with a difference of about 16\% for both specificity and sensitivity. In addition, our attempt to use the trachea information in complement to the audio modality did not lead to any improvement. A very similar conclusion can be drawn for our experiments with the contact microphone placed on the thorax (see Table \ref{tab:Thorax}), which were led on set B of the database with 10 subjects.

Our interpretation of this conclusion is the following. Although the contact microphone is more robust to ambient noise, it suffers from other acquisition perturbations, notably due to the fact that with such a sensor, it is difficult in some conditions (subject moving, etc) to maintain a good contact with the skin, leading to some degradation during the acquisition. Furthermore, even in noisy environments, the explosive phase captured by the audio microphone has a sufficiently high energy, keeping a high Signal-to-Noise Ratio during this timespan. Therefore, according to our experiments, using the contact microphone does not convey a sufficiently high amount of relevant information in complement with the audio signal. This lack of reliability of the contact microphone makes then us discard its possible use in a 24-hours ambulatory system.

\begin{table} 
  \begin{center}
\vspace{0cm}
    \begin{tabular}{c|c|c|}
 & \bf{Specificity (\%)} & \bf{Sensitivity (\%)} \\
\hline 
\hline 
\bf{Audio} & 89.97 & 89.85 \\
\hline 
\bf{Thorax} & 70.87 & 70.77 \\
\hline 
\bf{Audio + Thorax} & 88.68 & 89.04 \\
\hline 
    \end{tabular}
  \end{center}
  \vspace{-0.5cm}
  \caption{\emph{Results of cough event detection using the contact microphone placed on the thorax (set B of the database).}}
  \label{tab:Thorax}
\vspace{-.2cm}
\end{table}

\section{Comparison with the Karmelsonix System}\label{sec:KS}

The performance of the proposed approach is now compared to the Karmelsonix system, one of the few commercially available cough counters. Karmelsonix makes use of four sensors \cite{Vizel}: one audio microphone, two contact microphones (one at the trachea and one on the thorax), and a chest wall impedance belt. The acquisition system is also provided with an analysis software from which we here consider the cough detections.

In this experiment, both systems are compared on set B of the database (10 subjects) where parallel recordings are available. For the proposed approach, we consider only the audio modality trained on set A (22 subjects).

\begin{figure}[h]
  \begin{center}
   \includegraphics[width=8cm]{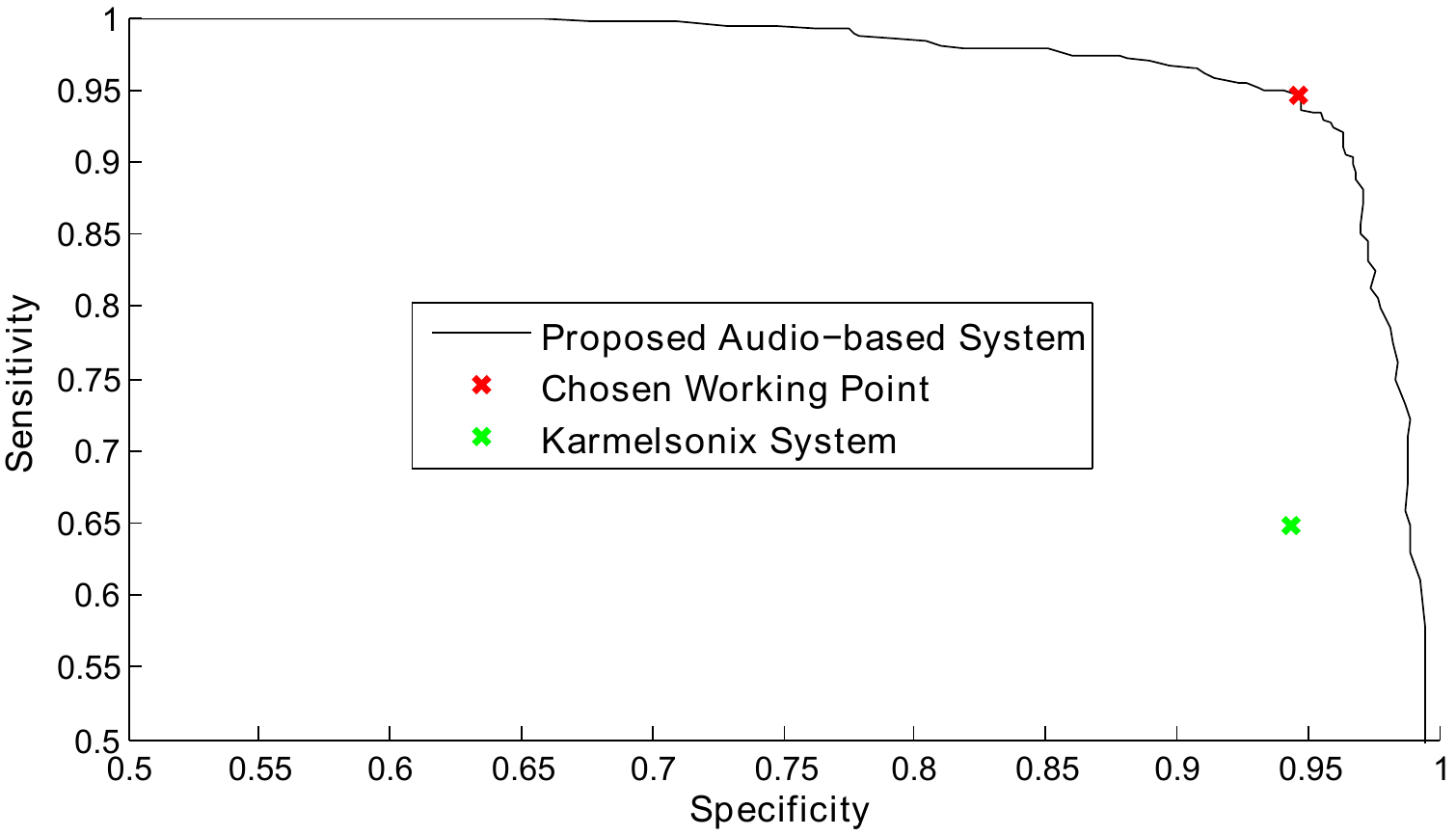}
  \end{center}
\vspace{-.5cm}
\caption{{\it Performance of the proposed audio-based cough detection and the Karmelsonix system in the specificity-sensitivity plane.}} 
\label{fig:ROC}
\vspace{-.2cm}
\end{figure}

Figure \ref{fig:ROC} displays the overall performance of the Karmelsonix systems and ours when the decision threshold decision is varied. It is noticed that although Karmelsonix gives good specificity results (few false alarms), its sensitivity is poor on average (about 65\%, indicating a high number of missed detections). Interestingly, the proposed approach is observed to provide, for a comparable specificity, much better sensitivity capabilities. As working point for the proposed technique, we fixed the decision threshold such that a compromise between misses and false alarms is found, as indicated in Figure \ref{fig:ROC}.

The detection results specific to each subject are presented in Table \ref{tab:CompKS}. It is noted that Karmelsonix suffers from a wide inter-subject variability. Indeed, while subjects 1, 8 and 10 have a specificity higher than 90\%, it turns out that the system misses almost all cough events for subjects 3 and 6. This drawback is the most important inconvenient of Karmelsonix. Note that this observation is in contradiction with the preliminary study carried out by Karmelsonix developers in \cite{Vizel} and where they report an overall sensitivity of 96\%. On the opposite, while the proposed approach achieved comparable specificity results, its sensitivity never went below 80\% and reached about 95\% on average.

To give an idea, these results can be compared to other similar studies, although they were not obtained on the same database. In \cite{Barry}, the HACC system based on the analysis of audio recordings achieved a specificity of 96\% and a sensitivity of 80\%. In \cite{Coyle}, it was reported that the commercialized LifeShirt system (using a microphone, a respiratory inductance plethysmography, and an accelerometer) gave a specificity and sensitivity of respectively 99.6\% and 78.1\%. Finally, the Leicester Cough Monitor was found in \cite{LCM} to have a specificity and sensitivity of respectively 99\% and 91\%.
 
\begin{table} 
  \begin{center}
\vspace{0cm}
    \begin{tabular}{c||c|c||c|c|}
    
 & \multicolumn{2}{c||}{\bf{Proposed}} & \multicolumn{2}{c|}{\bf{Karmelsonix}} \\
\hline 
\textbf{Subj.} & \textbf{Sens.(\%)} & \textbf{Spec.(\%)} & \textbf{Sens.(\%)} & \textbf{Spec.(\%)}\\
\hline 
1 & 100 & 98.6 & 93.1 & 100\\
\hline 
2 & 95.9 & 95.9 & 87.7 & 90.1\\
\hline 
3 & 86.1 & 95.4 & 0.0 & 100\\
\hline 
4 & 98.6 & 97.3 & 83.3 & 93.7\\
\hline 
5 & 98.6 & 91.0 & 61.4 & 86.0\\
\hline 
6 & 81.3 & 96.8 & 4.0 & 100\\
\hline 
7 & 98.6 & 100 & 68.1 & 92.4\\
\hline 
8 & 88.0 & 98.5 & 90.7 & 97.1\\
\hline 
9 & 100 & 80.9 & 70.4 & 96.2\\
\hline 
10 & 100 & 96.0 & 90.1 & 97.0\\
\hline 
\hline 
\textbf{Avg.} & \textbf{94.7} & \textbf{95.0} & \textbf{64.9} & \textbf{95.3}\\
\hline 
\textbf{STDV} & 6.9 & 5.5 & 34.8 & 4.7\\
\hline 
    \end{tabular}
  \end{center}
  \vspace{-0.5cm}
  \caption{\emph{Detail of the performance achieved by the proposed and Karmelsonix systems for the 10 subjects of set B.}}  
  \label{tab:CompKS}
\vspace{-.2cm}
\end{table}

Finally, Table \ref{tab:Confusion} highlights the sources of false alarms for the proposed audio-based technique. Interestingly, segments of speech and background noise are well discriminated by the proposed method, leading to only one false alarm over 58 minutes. The most confusing audio classes are respectively laugh (15.6\%), throat clearing (8.1\%) and forced expiration (3.3\%).

\begin{table} 
  \begin{center}
\vspace{0cm}
    \begin{tabular}{c|c|c|}
 & \bf{Amount} & \bf{\# Cough detected} \\
\hline 
\hline 
\bf{Cough} & 726 & 687 \\
\hline 
\bf{Expiration} & 90 & 3 \\
\hline 
\bf{Throat Clearing} & 148 & 12 \\
\hline 
\bf{Speech} & 474 sec & 1 \\
\hline 
\bf{Laugh} & 96 & 15 \\
\hline 
\bf{Background} & 3006 sec & 0 \\ 
\hline 
    \end{tabular}
  \end{center}
  \vspace{-0.5cm}
  \caption{\emph{Repartition of cough events detected by the proposed audio-based approach across the sound categories.}}
  \label{tab:Confusion}
\vspace{-.2cm}
\end{table}

\section{Conclusion}\label{sec:conclu}
This paper proposed a technique of automatic cough detection based on an audio microphone and the possible complementary use of a contact microphone. The proposed approach relies on a large variety of audio features, on an algorithm of feature selection based on Information Theory measures, and on the decomposition of the classification into two subtasks achieved by two separate ANNs. Our integration of the contact microphone (placed either on the thorax or the trachea) did not provide any improvement compared to the audio modality, as this sensor was subject to acquisition perturbations, leading to reliability issues. Finally, the proposed approach was shown to outperform the Karmelsonix system, one of the few commercially available solutions. Achieving a comparable specificity, the proposed method led to an absolute gain of 30\% in sensitivity. Our future works encompass the use of a HMM-based classifier and investigations about the potential use of other sensors such as an accelerometer, an ECG and a piezoelectric chest belt.

\section{Acknowledgements}
The project is supported by the Walloon Region (Grant WIST 3 COMPTOUX \# 1017071).

\eightpt
\bibliographystyle{IEEEtran}

\end{document}